%% file: 0_main.tex
\documentclass[sigconf,screen]{acmart}
\AtBeginDocument{%
  \providecommand\BibTeX{{%
    \normalfont B\kern-0.5em{\scshape i\kern-0.25em b}\kern-0.8em\TeX}}}

\usepackage{enumitem}
\usepackage{geometry}
\usepackage{listings}
\usepackage{xcolor}

\usepackage[T1]{fontenc}
\usepackage[utf8]{inputenc}

\usepackage{tcolorbox}
\usepackage{listings}
\usepackage{xcolor}

\definecolor{codebg}{RGB}{245,245,245}

\lstdefinestyle{promptstyle}{
  language=Python,
  basicstyle=\ttfamily\footnotesize,
  breaklines=true,
  backgroundcolor=\color{codebg},
  frame=none,
  columns=fullflexible,
  keepspaces=true,
  showstringspaces=false
}

\setcopyright{acmlicensed}
\copyrightyear{2026}
\acmYear{2026}
\acmDOI{XXXXXXX.XXXXXXX}

\acmConference[EASE 2026]{The 30th International Conference on Evaluation and Assessment in Software Engineering}{9–12 June, 2026}{Glasgow, Scotland, United Kingdom}
\acmISBN{978-1-4503-XXXX-X/18/06}

\begin{document}

\title[Making OpenAPI Documentation Agent-Ready with Hermes]{Making OpenAPI Documentation Agent-Ready: Detecting Documentation and REST Smells with a Multi-Agent LLM System}


\input{1_authors.tex}

\input{2_abstract.tex}
\renewcommand{\shortauthors}{Lima et al.}




\maketitle

\input{3_introduction}
\input{4_background}
\input{5_method}
\input{6_results}
\input{7_discussion}
\input{8_conclusion}

\begin{acks}
This paper is a result of the Research, Development \& Innovation Project named COPILOTO at Sidia Institute of Science and Technology sponsored by Samsung Eletrônica da Amazônia Ltda., using resources under terms of Federal Law No. 8.387/1991, by having its disclosure and publicity in accordance with art. 39 of Decree No. 10.521/2020.
\end{acks}

\bibliographystyle{ACM-Reference-Format}
\bibliography{reference}

\appendix
\section*{APPENDIX}
\input{9_apendiceA}









\end{document}

%% file: 1_authors.tex

\author{Rayfran Rocha Lima}
\email{rayfran.lima@sidia.com}
\orcid{0000-0002-9807-1266}
\affiliation{%
  \institution{Sidia Institute of Technology}
  \city{Manaus}
  \state{AM}
  \country{Brazil}
  }
  
\author{Davi G. Assuncao Pinheiro}
\email{davi.pinheiro-e@sidia.com}
\orcid{0009-0002-2733-8346}
\affiliation{
  \institution{Sidia Institute of Technology}
  \city{Manaus}
  \state{AM}
  \country{Brazil}
}

\author{Thiago Medeiros de Menezes}
\email{thiago.menezes@sidia.com}
\orcid{0000-0001-5154-4540}
\affiliation{
 \institution{Sidia Institute of Technology}
  \city{Manaus}
  \state{AM}
  \country{Brazil}
\email{thiago.menezes@sidia.com}
}

%% file: 2_abstract.tex
\begin{abstract}
The growing adoption of AI agents and the Model Context Protocol (MCP) has motivated organizations to expose existing REST APIs as agent-consumable tools. In our industrial context, this initiative targeted an ecosystem of 16 production APIs comprising approximately 600 endpoints. Although these APIs were stable and widely used within a microservice architecture, early proof-of-concept experiments revealed systematic failures in task planning, tool selection, and payload construction when accessed through MCP-based agents. Rather than attributing these failures to model limitations alone, we conducted an ecosystem-scale empirical assessment of the underlying OpenAPI documentation. We developed \textit{Hermes}, a multi-agent LLM-based system that detects documentation and REST-related smells at the endpoint level and generates explainable diagnostic reports. The large-scale evaluation identified 2,450 smells across 600 endpoints, with deficiencies present in all analyzed operations. Practitioner validation confirmed high agreement with the detected issues while also revealing contextual trade-offs in remediation decisions. The findings suggested that structural validity within microservice environments does not guarantee semantic readiness for agent-based consumption. Based on this evidence, the organization revised its adoption strategy, prioritizing selective endpoint adaptation, redefining documentation standards, and integrating automated documentation assessment into API governance workflows. This case illustrates how systematic artifact-level evaluation can function as a strategic decision-support mechanism, reducing technological risk and guiding evidence-based AI adoption in industrial software ecosystems.
\end{abstract}

%% file: 3_introduction.tex
\section{Introduction}
\label{sec:Introduction}

Application Programming Interfaces (APIs) play a central role in modern software ecosystems, acting as long-lived contracts that enable integration between heterogeneous systems and teams \cite{araujo2022openapi}. The OpenAPI Specification has become the de facto standard for documenting RESTful APIs, supporting automation activities such as client generation, testing, monitoring, and interactive documentation \cite{openapi20242}. Consequently, the quality of OpenAPI documentation directly impacts usability, maintainability, interoperability, and system reliability \cite{alzahrani2024software, robillard2017api}.

REST principles emphasize resource-oriented modeling, consistent use of HTTP methods, meaningful status codes, and explicit security definitions \cite{fielding2000rest, pautasso2014rest, masse2011rest}. Violations of these principles reduce predictability and reuse, affecting both human understanding and automated processing \cite{pautasso2014rest}. Prior studies report that OpenAPI specifications frequently contain incomplete, ambiguous, or inconsistent descriptions, especially when generated automatically or maintained under time constraints \cite{araujo2022openapi, coblenz2023qualitative}. Existing validation tools predominantly focus on syntactic correctness or structural conformance, offering limited support for assessing semantic documentation quality \cite{theunissen2022mapping}.

Recent advances in large language models (LLMs) have enabled the development of AI agents capable of autonomously interacting with external systems, relying heavily on contextual artifacts such as API specifications for planning, tool selection, parameter construction, and response interpretation \cite{du2024evaluating, white2023prompt}. Research highlights that agent performance is highly sensitive to the clarity, structure, and semantic richness of tool descriptions and contextual inputs \cite{bandlamudi2025framework, ni2025doc2agent, guo2025mcp}. 

The Model Context Protocol (MCP) has been introduced as a standardized mechanism for exposing APIs as agent-consumable tools \cite{anthropic2024mcp}, with OpenAPI specifications often serving as the primary knowledge source guiding agent reasoning. While studies demonstrate the feasibility of transforming OpenAPI specifications into tool-augmented agent environments \cite{mastouri2025making, ni2025doc2agent}, documentation quality is frequently assumed rather than systematically analyzed. Failures are typically detected at runtime—during tool invocation or parameter mismatch—without addressing whether deficiencies stem from the underlying documentation \cite{mastouri2025making, li2025we}.

Documentation smells extend the notion of code smells to textual artifacts, capturing recurring deficiencies such as vague descriptions, fragmented explanations, or mixed concerns \cite{khan2021docs, panichella2022test}. Although their impact on maintainability and onboarding has been documented, systematic detection of documentation smells in OpenAPI specifications remains underexplored. Moreover, limited work explicitly connects documentation quality to the reliability of agent-based API consumption.

Motivated by these gaps, we conducted an industrial feasibility assessment in the context of building an AI-agent-based automation ecosystem integrated through MCP servers. However, early proof-of-concept (PoC) experiments revealed systematic failures when these APIs were exposed as MCP tools for agent interaction.

Rather than attributing these limitations solely to model constraints, we hypothesized—consistent with prior literature on context sensitivity in LLM-based systems—that structural and semantic deficiencies in the OpenAPI documentation could emerge as a central limiting factor. The PoC functioned as a diagnostic probe reinforcing evidence that agent behavior is strongly influenced by the semantic clarity of tool descriptions.

To investigate documentation readiness at scale, we introduced \textit{Hermes}, a multi-agent LLM-based system designed as an evaluation instrument to detect documentation and REST-related smells at the endpoint level and generate explainable diagnostic reports. The objective was not to benchmark agent performance directly, but to assess whether the existing API ecosystem provided sufficient semantic clarity to support reliable agent-based consumption.

This paper reports the results of this empirical evaluation conducted in a real-world industrial setting. Our contributions:

\begin{itemize}
    \item An industrial-scale empirical assessment of documentation and REST-related quality issues.
    \item A practitioner-validated analysis of identified deficiencies, highlighting contextual trade-offs in remediation decisions.
    \item Evidence demonstrating how systematic documentation assessment informed strategic AI adoption decisions and reduced projected remediation effort in an industrial software ecosystem.
\end{itemize}

By presenting this case, we illustrate how artifact-level empirical evaluation can guide technology adoption, support cost-benefit analysis, and strengthen quality governance in AI-enabled industrial software engineering contexts.

%% file: 4_background.tex
\section{Background and Related Work}
\label{sec:background}

OpenAPI serves as a formal contract describing endpoints, parameters, request bodies, responses, and security mechanisms \cite{casas2021uses}. In microservice architectures, it frequently operates as a single source of truth for integration and orchestration tasks. However, prior empirical studies report that OpenAPI documentation often becomes semantically underspecified, particularly when automatically generated from source code or maintained under delivery pressure \cite{araujo2022openapi, robillard2017api}.

While static analysis tools detect syntactic and structural violations in OpenAPI specifications, they provide limited support for evaluating semantic expressiveness, clarity, and coherence \cite{theunissen2022mapping}. As a result, specifications may remain structurally valid yet insufficiently descriptive for external consumers or automated systems.

Documentation smells extend the concept of code smells to textual artifacts, capturing recurring quality deficiencies such as vague descriptions, fragmented explanations, and mixed concerns \cite{khan2021docs, alzahrani2024software}. While prior research has documented the influence of documentation quality on maintainability and onboarding, there is limited work explicitly connecting these deficiencies to the reliability of agent-based API consumption. Most prior work focuses on natural language documentation embedded in source code \cite{ni2025doc2agent}. Little research operationalizes documentation smells in structured API specifications or examines their implications for automated consumption by AI agents.

Recent advances in LLM-based agents enable automated interaction with external tools and APIs \cite{white2023prompt, du2024evaluating}. Frameworks such as MCP facilitate exposing APIs as agent-consumable tools \cite{anthropic2024mcp}. Research on tool-augmented LLMs and API invocation demonstrates that agents can successfully perform complex multi-step tasks when provided with sufficiently expressive tool descriptions and contextual guidance \cite{bandlamudi2025framework, ni2025doc2agent}.

However, existing approaches predominantly evaluate execution correctness or runtime performance \cite{guo2025mcp}. Documentation quality is often treated as an implicit prerequisite rather than an explicit object of empirical assessment. Failures are typically identified during execution without systematically tracing them to structural or semantic deficiencies in the specification \cite{mastouri2025making, lima2025ictai}.

This gap is particularly relevant in industrial contexts where APIs were not originally designed for agent-based consumption. Understanding the documentation readiness of an API ecosystem—before large-scale MCP conversion—becomes a critical feasibility criterion.

Although grounded in an industrial context, similar documentation challenges—such as incomplete, ambiguous, and inconsistent specifications—are also observed in public APIs and open ecosystems. These issues often reflect implicit knowledge dependencies, hindering understanding by new consumers~\cite{coblenz2023qualitative}.

Taken together, prior research emphasizes runtime validation and infrastructure scalability, but offers limited support for preventive, documentation-centric evaluation of OpenAPI specifications at ecosystem scale. This study addresses that gap through an industrial case study focused on artifact-level assessment as a foundation for evidence-based AI adoption. 

%% file: 5_method.tex
\section{Industrial Context and Feasibility Assessment Design}
\label{sec:industrial_context}

\subsection{Research Design and Case Study Framing}

This study follows an industrial case study design in accordance with established guidelines for case study research in software engineering \cite{runeson2009case}. The objective is exploratory and evaluative: to investigate whether documentation readiness constitutes a limiting factor for large-scale AI-agent adoption in an existing industrial API ecosystem.

The case is embedded within a single industrial organization, with the OpenAPI endpoint defined as the primary unit of analysis. The ecosystem of active APIs used by the company defines the boundary of the case.

The study does not aim to establish causal inference between documentation quality and agent performance. Instead, it evaluates documentation readiness as a feasibility criterion for technology adoption. Multiple sources of evidence were collected and analyzed, including: (i) proof-of-concept experimentation, (ii) large-scale automated smell detection, (iii) expert validation of a representative subset, and (iv) practitioner review of diagnostic reports. This triangulation increases construct validity and strengthens the robustness of the findings.

\subsection{Organizational Context}

The study was conducted within an industrial R\&D institute operating as part of a global network focused on embedded systems development. The organization maintains a microservice-based architecture supporting project management, product configuration, integration processes, and interaction with external platforms such as issue tracking systems.

At the time of the study, the organization maintained 16 production APIs comprising approximately 600 endpoints. These APIs were actively used for front-end integration and service-to-service communication and were considered stable within their intended microservice context. However, they were not originally designed with external reuse or autonomous agent consumption in mind.

Driven by advances in AI agents and the emergence of the Model Context Protocol (MCP), the organization initiated an internal initiative to build an AI-based automation ecosystem. The strategy initially assumed that existing OpenAPI specifications could be directly converted into MCP servers, thereby exposing endpoints as agent-consumable tools.

\subsection{Proof-of-Concept as Diagnostic Probe}

Prior literature on LLM-based tool invocation and API-mediated agents indicates that agent performance is highly sensitive to the semantic clarity and expressiveness of tool descriptions and contextual artifacts \cite{bandlamudi2025framework, ni2025doc2agent}. Our proof-of-concept (PoC) did not seek to establish this sensitivity as a novel phenomenon, but to assess whether similar effects manifested within our industrial ecosystem.

To this end, we conducted a controlled diagnostic experiment using locally deployed large language models integrated with MCP-generated servers derived from OpenAPI specifications.

Two MCP servers were constructed. The first was generated directly from a FastAPI application containing 10 endpoints, preserving the original OpenAPI documentation without refinement. The second server exposed the same business logic and endpoints but incorporated enriched documentation through detailed docstrings explicitly describing tool objectives, expected inputs, security requirements, output structures, and operational semantics.

Ten representative automation tasks were defined. Eight required coordination between multiple tools, and two involved single-tool interaction. These tasks reflected realistic project management scenarios.

When executing tasks through agents interacting with the first MCP server (original documentation), approximately 70\% of tasks failed during the planning phase. Only 10\% of expected endpoints were correctly identified, and none were successfully executed end-to-end due to payload construction or authentication errors.

In contrast, with enriched documentation, correct execution plans were generated in 90\% of tasks, 80\% of endpoints were correctly identified, and all identified endpoints were successfully executed.

Because both servers exposed identical business logic, the most salient experimental difference was the documentation enrichment. These findings aligned with prior research suggesting that agent reasoning is highly dependent on contextual clarity. The PoC therefore functioned as a diagnostic probe, reinforcing literature-based expectations and motivating a systematic, documentation-centric feasibility assessment at the ecosystem scale.

The PoC was designed as an exploratory diagnostic rather than a controlled experiment. Each task was executed once per configuration, without repeated runs or statistical treatment. Given the stochastic nature of LLM outputs, no explicit control of randomness (e.g., temperature or seed variation) was applied. Therefore, the results should be interpreted as indicative evidence of systemic behavior rather than definitive performance measurements. Therefore, results should not be interpreted as statistically generalizable.

\subsection{Rationale for Documentation-Centric Assessment}

Rather than immediately scaling MCP conversion, the organization decided to conduct a structured feasibility assessment focused on documentation readiness.

Preliminary inspection suggested that fields such as \textit{summary} and \textit{description} were frequently sparse, inconsistent, or oriented toward implicit team knowledge. Endpoint semantics were occasionally ambiguous, HTTP methods diverged from REST conventions, and parameter schemas relied on generic structures with limited explanatory detail.

While such characteristics did not hinder internal human developers, they posed potential risks for autonomous agents relying exclusively on OpenAPI metadata. Given the scale of the ecosystem (600 endpoints), manual inspection was impractical. A comprehensive refactoring effort was estimated at approximately 385 engineering hours, requiring a more evidence-based decision strategy.

\subsection{Feasibility Assessment Design}
\label{feasibility}

To operationalize documentation readiness evaluation at scale, we developed \textit{Hermes} as an assessment instrument. Hermes detects documentation and REST-related smells at the endpoint level and generates structured, explainable reports.

Building upon the work of Khan et al.~\cite{khan2021docs}, this study interprets \textbf{documentation smells} as indicators of insufficient semantic expressiveness that may hinder both human understanding and autonomous agent consumption. Four smell categories are considered: \textbf{LAZY:} Superficial or generic documentation, including short summaries, vague descriptions, undocumented parameters, or generic response messages. \textbf{BLOATED:} Excessively verbose descriptions with limited informational value. \textbf{TANGLED:} Documentation mixing unrelated concerns (e.g., business logic, security, error handling) within the same textual fragment. \textbf{FRAGMENTED:} Essential information dispersed across disconnected sections without explicit linkage.


Grounded in established REST design principles~\cite{fielding2000rest, masse2011rest, coblenz2023qualitative}, \textbf{REST-related smells} capture recurring design inconsistencies that reduce predictability and interoperability. Five categories are considered: \textbf{PATH:} Action-oriented or inconsistently named URIs that do not represent resources. \textbf{METHOD:} Inappropriate or semantically inconsistent use of HTTP methods. \textbf{INPUT:} Weakly specified parameters or request bodies lacking semantic clarification. \textbf{RESPONSE:} Inconsistent or insufficiently described response schemas, status codes, or error handling. \textbf{SECURITY:} Missing or unclear authentication and authorization definitions.


Hermes is designed to evaluate documentation readiness of OpenAPI specifications at the endpoint level. Given the URL of an OpenAPI JSON file, the system enables interactive exploration and automated smell detection for individual API operations.

Rather than analyzing the specification as a monolithic artifact, Hermes adopts an \textit{endpoint-centric} strategy: each operation (defined by the combination of HTTP method and path) is isolated and evaluated independently. For a selected endpoint, the system generates a reduced OpenAPI representation containing only the relevant elements (method, path, summaries, descriptions, parameters, request body, responses, schemas, and security definitions). An illustrative excerpt of the reduced OpenAPI representation used in the analysis is provided in Appendix B.

This reduction serves two purposes. First, it decreases token consumption for large language models, addressing context-window limitations. Second, it improves analytical focus by removing unrelated endpoints and documentation fragments that could introduce ambiguity. The resulting artifact preserves all semantic information necessary for smell detection while enabling scalable analysis across large ecosystems.

Hermes follows a multi-agent architecture. A central Smell Detector Agent orchestrates the workflow and consolidates results, while specialized agents analyze specific smell categories. Documentation Smell Agents evaluate textual deficiencies (e.g., LAZY, BLOATED, TANGLED, FRAGMENTED), and REST Smell Agents evaluate design-related inconsistencies (e.g., PATH, METHOD, INPUT, RESPONSE, SECURITY). The prompt template used by the specialized agents, including the definition and classification rules for the Lazy documentation smell, is provided in Appendix A.

Each specialized agent operates based on a structured prompt that encodes the definition of the target smell, along with classification criteria and illustrative examples following a few-shot prompting strategy. All agents receive the same reduced endpoint-specific OpenAPI representation but analyze it from distinct perspectives according to their assigned smell category. The prompts guide agents to produce structured, explainable outputs, including justification of detected issues and suggested improvements, supporting consistency and interpretability across analyses.

All agents receive the same reduced endpoint representation and apply structured prompting based on the smell taxonomy. Their outputs are aggregated into a unified, explainable diagnostic report. This modular design supports extensibility and enables consistent, scalable documentation assessment across the entire API ecosystem.

Since Hermes relies on large language models (LLMs) for semantic smell detection, selecting a model with reliable multi-label classification performance was necessary before conducting the large-scale ecosystem assessment.

A representative subset of 60 endpoints (10\% of the 600 production endpoints) was randomly selected to construct a gold standard dataset. Two domain experts manually annotated documentation and REST-related smells according to the taxonomy defined in Section~3. Disagreements were resolved through discussion to produce consolidated reference labels.

Smell detection was formulated as a multi-label classification task, and model performance was evaluated using standard agreement metrics, including Jaccard similarity, F1 scores (micro and macro), and Hamming loss. These metrics collectively capture overlap with expert annotations, balance across frequent and infrequent smell categories, and prediction error rate.

Seven LLMs were evaluated under identical prompting conditions to ensure comparability. As shown in Table~\ref{tab:rq1_global_metrics}, \textit{gpt-oss:120b} achieved the highest agreement with the expert-annotated gold standard (Jaccard = 0.85; F1$_{micro}$ = 0.92; F1$_{macro}$ = 0.73) and the lowest Hamming loss (0.07), demonstrating more stable multi-label behavior across smell categories. Cardinality difference (column Cardin. Diff) presents the deviation between predicted and actual number of labels per instance.

\begin{table}[htp]
\centering
\caption{Overall accuracy metrics for smell detection per LLM}
\label{tab:rq1_global_metrics}
\small
\begin{tabular}{p{2.3cm}p{.8cm}p{.8cm}p{.8cm}p{.8cm}p{.8cm}}
\hline
Model & Jaccard & F1$_{micro}$ & F1$_{macro}$ & Ham-ming & Cardin. Diff \\
\hline
\textbf{gpt-oss:120b} & \textbf{0.85} & \textbf{0.92} & \textbf{0.73} & \textbf{0.07} & \textbf{-0.53} \\
gemma3:27b & 0.53 & 0.70 & 0.50 & 0.37 & +1.83 \\
qwen2.5vl:72b & 0.67 & 0.80 & 0.49 & 0.20 & -0.47 \\
deepseek-r1:70b & 0.69 & 0.81 & 0.48 & 0.16 & -1.22 \\
llama3.2-vision:90b & 0.62 & 0.77 & 0.46 & 0.21 & -1.17 \\
llama4:16x17b & 0.51 & 0.68 & 0.44 & 0.29 & -1.18 \\
qwen3-coder:30b & 0.53 & 0.70 & 0.38 & 0.25 & -1.72 \\
\hline
\end{tabular}
\end{table}

Given its superior alignment with expert judgment and balanced prediction behavior, \textit{gpt-oss:120b} was selected as the operational model for the full ecosystem assessment. The purpose of this evaluation was not to benchmark LLMs broadly, but to ensure analytical reliability before applying Hermes to all 600 industrial endpoints.

This selection step increases confidence that the large-scale documentation readiness results reflect ecosystem characteristics rather than model-specific prediction bias.

The feasibility assessment comprised four coordinated activities:

\begin{itemize}
    \item \textbf{Gold Standard Definition:} Manual annotation of a representative 10\% subset (60 endpoints) by domain experts to evaluate model reliability.
    \item \textbf{Practitioner Evaluation:} Structured review of selected diagnostic reports by API authors and maintainers.
    \item \textbf{Large-Scale Technical Assessment:} Automated smell detection across all 600 endpoints.
    \item \textbf{Strategic Cost Estimation:} Comparative estimation of remediation effort under full-ecosystem refactoring versus selective endpoint adaptation.
\end{itemize}

Data analysis combined quantitative ecosystem-level metrics (e.g., smell frequency and distribution), multi-label classification accuracy measures, qualitative practitioner feedback, and cost comparison analysis. This combination of quantitative and qualitative evidence supports analytical triangulation and strengthens internal validity.

\subsection{Industrial Evaluation Questions}

To structure the analysis, we formulated three Industrial Evaluation Questions (IEQs):

\begin{itemize}
    \item \textbf{IEQ1 (Documentation Readiness):} To what extent does the OpenAPI ecosystem exhibit documentation and REST-related deficiencies that may compromise reliable agent-based consumption?
    
    \item \textbf{IEQ2 (Practitioner Validation):} To what extent do practitioners agree with identified deficiencies, and how do contextual constraints shape remediation decisions?
    
    \item \textbf{IEQ3 (Strategic Feasibility):} What are the feasibility and cost implications of converting the API ecosystem into MCP-based agent tools under current documentation conditions?
\end{itemize}

These questions frame the study as an industrial feasibility evaluation rather than a benchmark of agent performance, aligning the investigation with decision-making needs in technology adoption contexts.

%% file: 6_results.tex
\section{Results}
\label{sec:results}

This section presents the findings structured around the Industrial Evaluation Questions (IEQs), emphasizing documentation readiness, practitioner validation, and strategic feasibility implications.

\subsection{IEQ1 – Documentation Readiness}
\label{sec:results-ieq1}

IEQ1 investigates the extent to which the existing OpenAPI ecosystem exhibits documentation and REST-related deficiencies that could compromise reliable agent-based consumption.

Following the model selection process described in Section~\ref{feasibility}, the selected LLM (\textit{gpt-oss:120b}) was applied to the full ecosystem comprising 600 production endpoints distributed across 16 APIs. All endpoints were active, validated, and operational within the organization's microservice architecture at the time of analysis.

\paragraph{Overall Smell Distribution.}

The automated assessment performed by \textit{Hermes} identified a total of 2,450 documentation and REST-related smells across the 600 analyzed endpoints. At least one smell was detected in every endpoint, resulting in an average of 4.08 smells per endpoint. Rather than indicating critical design failure, this distribution suggests that documentation and design deficiencies were pervasive and systematically embedded in the ecosystem. The findings point to recurring patterns of semantic underspecification and implicit assumptions that, while manageable within human-centered workflows, may introduce structural risks under autonomous agent consumption.

Table~\ref{tab:ieq1_smell_distribution} presents the frequency and percentage distribution of detected smells per category. 

\begin{table}[htp]
\centering
\caption{Distribution of detected documentation and REST-related smells.}
\label{tab:ieq1_smell_distribution}
\small
\begin{tabular}{lcc}
\hline
\textbf{Smell Category} & \textbf{Frequency} & \textbf{\% of Endpoints} \\
\hline
Response & 600 & 100\% \\
Lazy & 540 & 90\% \\
Input & 530 & 88\% \\
Security & 410 & 68\% \\
Path\_and\_Method & 320 & 53\% \\
Tangled & 30 & 5\% \\
Bloated & 10 & 2\% \\
Fragmented & 10 & 2\% \\
\hline
\end{tabular}
\end{table}

Response-related smells were identified in all endpoints, indicating that although response schemas were technically defined, they frequently lacked explicit semantic clarification regarding returned structures, success/error interpretation, or usage semantics. An illustrative example of a complete Hermes diagnostic report and its corresponding OpenAPI specification excerpt is provided in Appendices B and C.

A recurring structural pattern observed in several endpoints involved generic DTO-based responses structured as { status, data }, where the data field could contain arbitrarily shaped JSON payloads without explicit schema constraints. While such patterns were feasible and sufficient for the original developers during implementation, they introduced semantic opacity over time. In some cases, even the original endpoint authors were unable to precisely describe the possible contents of the data field without revisiting the source code. This finding illustrates that response-related smells did not merely reflect documentation sparsity, but exposed deeper issues of implicit knowledge embedded in implementation rather than in the OpenAPI specification.

Similarly, \textit{Lazy} (90\%) and \textit{Input} (88\%) smells were highly prevalent, revealing widespread sparsity or ambiguity in endpoint summaries, descriptions, and parameter documentation.

Security-related smells (68\%) suggest that authentication and authorization mechanisms, while structurally specified, often lacked operational guidance or contextual explanation. Additionally, 53\% of endpoints exhibited \textit{Path\_and\_Method} inconsistencies, reflecting deviations from REST design conventions such as action-oriented URIs or non-idiomatic HTTP method usage~\cite{gowda2024best}.

In contrast, textual excess smells such as \textit{Bloated} (2\%) and \textit{Tangled} (5\%) were comparatively rare. This distribution aligns with the minimalistic documentation style observed in the organization, where documentation tended to be sparse rather than overly verbose.

\paragraph{Recurring Evidence Patterns.}

These patterns are further illustrated through a concrete example combining specification and analysis report in Appendices B and C. Table~\ref{tab:rq1_evidence_patterns} summarizes the main recurring evidence patterns observed in the reports, capturing how smells typically manifested in the analyzed OpenAPI specifications.

\begin{table*}[t]
\centering
\caption{Recurring evidence patterns supporting smell identification in Hermes reports.}
\label{tab:rq1_evidence_patterns}
\small
\begin{tabular}{p{2.0cm}p{15.1cm}}
\hline
\textbf{Smell} & \textbf{Main recurring evidence patterns observed in Hermes reports} \\
\hline
\textit{Lazy} &
Very short or generic \textit{summary} (e.g., ``Get data''), absent or redundant \textit{description}, and occasional semantic mismatch between \textit{summary} and \textit{description}. Limited contextual explanation of the endpoint purpose and lack of illustrative examples or constraints, requiring external knowledge to interpret intent. \\
\hline
\textit{Input} &
Parameters specified only by type (e.g., \texttt{string}, \texttt{integer}) with little or no semantic description; ambiguous parameter names (abbreviations or internal acronyms) without explanation; missing information about valid ranges, accepted formats, or behavioral impact; required parameters without functional rationale. \\
\hline
\textit{Response} &
Presence of response schemas (often via \texttt{\$ref}) with minimal explanatory text; generic response descriptions (e.g., ``Successful Response''); limited semantic clarification of payload meaning across success and error cases; insufficient guidance on how consumers should interpret returned structures. \\
\hline
\textit{Security} &
Security schemes technically defined (e.g., Bearer token) but lacking operational guidance (how to obtain credentials, required scopes/permissions, or access constraints); security treated as implicit organizational knowledge rather than documented usage instructions. \\
\hline
\textit{Path\_and\_Method} &
Non-idiomatic or inconsistent HTTP method usage (e.g., using \texttt{POST} for updates that would typically be modeled as \texttt{PUT} or \texttt{PATCH}); recurrent action-oriented URIs with verbs embedded in the path (e.g., \texttt{/getUsers}, \texttt{/updateStatus}), rather than resource-oriented naming; long or semantically opaque paths containing internal terms or acronyms without clarification. \\
\hline
\textit{Bloated} / \textit{Tangled} (rare) &
Textual fields (\textit{summary}/\textit{description}) were generally short and low-density, reducing opportunities for excessive verbosity (\textit{Bloated}) or mixing multiple concerns within the same documentation fragment (\textit{Tangled}). Their scarcity is consistent with minimalistic documentation practices rather than evidence of consistently high-quality narratives. \\
\hline
\textit{Excessive Structured} (absent) &
The agent was configured to distinguish normative OpenAPI structure (schemas and references under \textit{components}) from cases where excessively structured content is embedded in natural language fields. The smell would apply if class-like definitions, nested structures, or formal specifications were written inside \textit{summary} or \textit{description}; however, no such cases were identified in the reports. \\
\hline
\textit{Fragmented} (absent) &
All referenced schemas/components were defined within the same OpenAPI document (typically under \textit{components}), with no broken or missing references. Fragmentation would be identified if endpoints referenced schemas/components not present in the specification, resulting in incomplete documentation; this condition was not observed in the analyzed reports. \\
\hline
\end{tabular}
\end{table*}

These evidence patterns help clarify why smell detection is more reliable for categories grounded in explicit structural elements of OpenAPI specifications (e.g., \textit{Input}, \textit{Response}, and \textit{Security}) and more dependent on contextual writing quality for smells centered on natural language fields (e.g., \textit{Lazy}). 

In the studied organizational context, endpoint \textit{summaries} and \textit{descriptions} were often short, generic, and occasionally semantically inconsistent. Documentation was primarily produced to support internal microservice integration and relied heavily on tacit knowledge shared among teams. While such practices did not hinder human developers familiar with the system, they reduced interpretability for agents relying exclusively on specification metadata.

\paragraph{Reliability of Large-Scale Assessment.}

As reported in Section~\ref{feasibility}, the selected LLM achieved strong agreement with expert annotations on a representative subset of 60 endpoints (Jaccard = 0.85; F1$_{micro}$ = 0.92; Hamming Loss = 0.07). This level of alignment provides confidence that the large-scale ecosystem results reflect structural documentation characteristics rather than model-specific prediction bias.

\paragraph{Implications for Agent Readiness.}

From a documentation readiness perspective, these findings indicate that although the APIs were structurally valid and operational within a microservice environment, their specifications lacked the semantic richness and explicitness required for autonomous agent consumption. 

Structural correctness did not imply agent-readiness. High prevalence of input, response, and description-related smells suggests that the ecosystem was optimized for human developers operating within shared organizational context, rather than for external or autonomous consumers requiring explicit and self-contained documentation cues.

Therefore, IEQ1 reveals that documentation and REST-related deficiencies were systemic across the ecosystem and represent a structural risk for reliable MCP-based agent integration under the existing documentation conditions.

\subsection{IEQ2 – Practitioner Validation and Contextual Constraints}
\label{sec:results-ieq2}

IEQ2 examines practitioner agreement with the identified deficiencies and explores how contextual constraints influence remediation decisions. Responses were collected using a Likert-scale questionnaire and analyzed through descriptive statistics, complemented by qualitative interpretation of practitioner feedback.

Using the same diagnostic reports employed during the LLM evaluation against the gold standard (a sample of 60 endpoints), we collected feedback from 24 developers, including API authors and maintainers, through a structured questionnaire.

Overall, practitioners demonstrated high agreement (80\% strongly agree, 7\% agree) with smells grounded in explicit and verifiable documentation elements, particularly those related to incomplete parameter descriptions, insufficient response explanations, and missing or unclear security definitions. These issues were widely perceived as objective documentation deficiencies.

However, disagreement (30\% disagree) emerged in cases where smell classification intersected with contextual trade-offs. The \textit{Lazy} smell — characterized by overly brief or generic summaries — was frequently contested. Several practitioners initially justified concise descriptions as sufficient within stable, internally shared microservice environments. When prompted to consider external consumers or AI agents relying solely on the OpenAPI specification, participants more readily acknowledged the limitations of such documentation practices.

Similar patterns were observed for REST-related smells such as non-idiomatic method usage or action-oriented URIs (20\% disagree, 3\% strong disagree). While deviations from REST conventions were recognized, they were sometimes justified by legacy constraints or organizational standards (32\% tried to justify). This highlights the sociotechnical nature of documentation quality: deficiencies are not merely technical artifacts but also reflections of inherited practices and implicit assumptions about API consumers.

Importantly, practitioners reported a learning effect during the evaluation process. Reviewing diagnostic reports exposed implicit documentation assumptions and increased awareness of REST design principles. This suggests that documentation assessment can function not only as a quality control mechanism but also as a reflective organizational artifact.

\subsection{IEQ3 – Strategic Feasibility and Cost Implications}
\label{sec:results-ieq3}

IEQ3 addresses the feasibility and cost implications of converting the existing API ecosystem into MCP-based agent tools under current documentation conditions. The selected 42 endpoints were directly derived from the functional requirements of the 18 defined automation scenarios, ensuring alignment between documentation assessment and practical use cases.

The initial proof-of-concept experiments revealed that approximately 70\% of evaluated automation scenarios failed before producing a valid execution plan. In other cases, semantically incorrect tools were selected due to ambiguous endpoint descriptions, leading to invalid payload construction or response misinterpretation. These failures were not attributable to API instability, but to insufficient semantic cues in the OpenAPI specifications guiding agent reasoning.

Based on the documentation assessment results, converting all 600 endpoints into MCP servers without prior remediation would likely propagate these limitations at scale. A preliminary estimation indicated that fully refactoring documentation across the entire ecosystem would require approximately 385 hours of coordinated engineering effort. In contrast, prioritizing the 42 endpoints directly required for the targeted 18 automation tasks reduced the estimated effort to approximately 42 hours, representing a substantial reduction (89\%) in required engineering effort.

These findings informed a strategic revision of the adoption plan. Instead of large-scale MCP conversion, the organization adopted a staged approach: (i) selective adaptation of endpoints relevant to defined automation scenarios, (ii) refinement of documentation standards aligned with agent-consumption requirements, and (iii) integration of \textit{Hermes} into the endpoint review process as a preventive assessment mechanism.

From a feasibility standpoint, the evaluation demonstrated that the primary limiting factor was not the underlying AI models or MCP infrastructure, but the documentation readiness of the API ecosystem. The assessment therefore functioned as a risk identification mechanism, enabling evidence-based redirection of investment and preventing large-scale rework under inadequate documentation conditions.

%% file: 7_discussion.tex
\section{Discussion}
\label{sec:discussion}

This industrial case highlights a structural lesson for organizations pursuing AI-agent integration: the readiness of underlying software artifacts can be a more decisive feasibility constraint than model capability or infrastructure maturity.

Although the studied API ecosystem was operationally stable within a microservice architecture, the transition to autonomous agent consumption exposed latent semantic fragilities. Documentation practices sufficient for human developers operating under shared contextual knowledge proved inadequate for agents relying exclusively on explicit OpenAPI metadata. In this sense, documentation deficiencies functioned as a form of architectural debt—dormant in human-mediated workflows, yet activated under autonomous reasoning.

The proof-of-concept experiments acted as a stress test for the ecosystem. Rather than revealing API instability, they exposed misalignment between the semantic richness required for agent planning and the level of explicitness in the specifications. The primary bottleneck was not AI capability, but artifact maturity. By reframing the problem as a documentation-readiness issue, the organization avoided misdirected investment in increasingly complex agent configurations.

The large-scale assessment further demonstrated that structural validity does not imply agent-readiness. OpenAPI specifications may be syntactically correct and operational within internal environments while lacking the semantic clarity required for reliable autonomous consumption. This gap underscores the need to treat documentation not merely as a descriptive artifact, but as an executable cognitive contract shaping machine reasoning.

From a strategic perspective, the evaluation enabled an evidence-based revision of the AI adoption plan. Instead of full-ecosystem MCP conversion, the organization adopted a scenario-driven adaptation strategy focused on endpoints supporting defined automation use cases. This shift reduced projected remediation effort and mitigated the risk of propagating systemic failures. The key insight is methodological: ecosystem-level readiness assessment should precede large-scale technological rollout.

An additional outcome was the transformation of Hermes from a diagnostic instrument into a governance mechanism. By integrating automated documentation assessment into endpoint review workflows, the organization institutionalized preventive quality control, making documentation readiness an explicit architectural concern. This illustrates how artifact-level evaluation can influence organizational practices beyond the initial technical objective.

More broadly, this case suggests that AI-agent adoption should be understood as a sociotechnical transition rather than a purely technological upgrade. Documentation smells and REST inconsistencies were often rooted in legacy conventions, implicit assumptions, and locally optimized design habits. The introduction of agents surfaced these assumptions and enabled a reflective space for revisiting documentation standards.

For industrial environments, three practical implications emerge. First, documentation readiness should be evaluated before exposing APIs as agent-consumable tools, as structural validity may conceal latent semantic deficiencies. Second, conformance to OpenAPI or REST principles does not guarantee semantic sufficiency, since agents depend on explicit, self-contained descriptions. Third, when documentation maturity is uneven, a selective, use-case-driven adaptation strategy may be more sustainable than full-ecosystem conversion.

Ultimately, this study reinforces a central EASE principle: disciplined evaluation must accompany technological enthusiasm. Evidence-based assessment of foundational software artifacts is essential to reduce strategic risk, optimize investment decisions, and enable sustainable integration of intelligent agents.

This case demonstrates that artifact-level readiness assessment can function as a gatekeeping mechanism in AI adoption roadmaps, enabling identification of structural limitations before large-scale deployment.

Although grounded in an industrial setting, these findings are expected to generalize to public API ecosystems and open platforms, where implicit knowledge assumptions are less viable and documentation must support a broader range of consumers, including both human developers and AI agents.

\section{Limitations}
\label{sec:limitations}

This study was conducted within a single industrial R\&D organization operating a microservice-based ecosystem composed of internally consumed APIs. Consequently, the specific distribution, prevalence, and types of documentation deficiencies observed may differ in public API platforms, open ecosystems, or organizations with distinct documentation cultures and governance practices. Although the quantitative patterns reported here may vary across contexts, the central insight—that documentation readiness influences the feasibility of agent-based API consumption—is likely transferable to environments characterized by legacy constraints, implicit knowledge dependencies, and documentation practices optimized for human-centric integration.

The assessment relied on large language models (LLMs) to perform semantic smell detection across OpenAPI specifications. While multiple models were comparatively evaluated under standardized prompting conditions to strengthen analytical reliability, some sensitivity to model configuration, inference dynamics, and prompt interpretation may remain. Model selection was additionally constrained to those capable of operating within the organization’s internal infrastructure. This restriction avoided transmitting proprietary API specifications to external servers and ensured compliance with internal security and confidentiality policies. Although this constraint may limit comparability with externally hosted frontier models, the strong agreement observed between the selected model and expert annotations increases confidence that the reported patterns reflect ecosystem characteristics rather than isolated model behavior.

The proof-of-concept component examined a limited set of representative automation scenarios rather than exhaustively evaluating all possible agent interactions. As such, reported failure rates should be interpreted as indicative rather than definitive performance benchmarks. Similarly, remediation effort estimates (e.g., 385 versus 42 engineering hours) represent planning-level approximations derived from internal cost projections. While absolute values may vary under different organizational conditions, the relative difference between full-ecosystem refactoring and selective adaptation remains strategically informative.

Finally, the evaluation reflects a specific technological configuration of MCP tooling and LLM capabilities at the time of analysis. Advances in agent reasoning, tool orchestration, or automated schema interpretation may reduce certain documentation-related limitations in future systems. However, relying exclusively on increasingly capable models to compensate for semantically ambiguous or underspecified API contracts may introduce additional operational and governance risks. For this reason, documentation quality is framed in this study as a structural readiness concern rather than a transient tooling limitation.

%% file: 8_conclusion.tex
\section{Conclusion}
\label{sec:conclusion}

This paper reported an industrial feasibility assessment conducted prior to large-scale adoption of AI-agent-based automation through MCP integration. Although the organization maintained a stable and operational API ecosystem composed of 16 production APIs and approximately 600 endpoints, the empirical evaluation suggested that structural and semantic deficiencies in the OpenAPI documentation may have constrained its readiness for autonomous agent consumption under the evaluated conditions.

Consistent with prior work, our findings reinforce that documentation quality constitutes a foundational dependency for reliable agent-based workflows. Structural validity and operational stability within microservice environments did not imply semantic sufficiency for AI-driven tool invocation. By systematically analyzing documentation and REST-related smells at scale, and validating findings with practitioners, the organization identified structural risks that would likely have compromised large-scale MCP conversion.

Rather than abandoning the strategic direction of AI-agent integration, the evidence supported a revised adoption strategy. The organization transitioned from full-ecosystem conversion to a selective, scenario-driven approach, prioritized documentation remediation where necessary, and integrated automated documentation assessment into its API governance process. This shift reduced projected remediation effort, mitigated infrastructure risk, and embedded documentation readiness as an explicit architectural concern.

Beyond the specific industrial context, this case illustrates the broader importance of evidence-based evaluation in technology adoption decisions. In environments characterized by rapid AI innovation, disciplined assessment of underlying software artifacts remains essential to ensure that enthusiasm for new paradigms is supported by structural readiness and sustainable governance practices.

By positioning documentation quality as a prerequisite rather than a byproduct of AI-agent integration, this study contributes to raising the profile of evaluation and assessment as core practices in industrial software engineering. These findings position documentation quality not as a secondary artifact, but as a foundational prerequisite for reliable AI-driven software ecosystems.

Future work may extend this evaluation in three directions. First, longitudinal studies could measure how documentation governance practices evolve after integrating preventive assessment mechanisms. Second, controlled experiments may quantify the impact of documentation remediation on agent performance metrics. Third, replication in different industrial domains and API ecosystems would strengthen the external validity of the findings.

%% file: 9_apendiceA.tex
\section{Example of Prompt Used by Specialized Agents}
\label{app:prompts}

\begin{tcolorbox}[
  title={Box A.1: Prompt Template for the Lazy Documentation Smell Agent},
  colback=white,
  colframe=black,
  boxrule=0.6pt
]
\begin{lstlisting}[style=promptstyle]
Instructions for the Specialist Agent

Role:
You are an expert in identifying "Lazy" documentation smells in API documentation.

Smell Definition (Lazy):
An API method should be classified as Lazy when its documentation is incomplete, vague, or generic, failing to provide essential information about the method's purpose, behavior, parameters, return values, or possible exceptions. This includes missing examples or unclear explanations that hinder correct understanding and usage.

This smell typically occurs when:
{lazy_examples}

Task:
Analyze the complete OpenAPI method definition provided below and identify whether the endpoint exhibits Lazy documentation.

OpenAPI Specification:
{openapi_json}

For each endpoint identified, you must classify the smell, justify the classification, and propose improvement actions.

Classification Rules:
- Analyze ONLY the method summary and description.
- Return ONLY a valid JSON object containing the endpoints with Lazy smells.
- Endpoint format: "GET:/users", "POST:/orders", etc.
- If no Lazy smell is found, return an empty JSON object.
- Do NOT include any text before or after the JSON output.

Explanation and Improvement Rules:
- Analyze ONLY the method summary and description.
- Provide explanations with a minimum length of 120 characters per section.
- Each explanation MUST contain exactly two sections:

1. "Justification and evidence of the smell:"
   - Present evidence of incomplete or vague documentation.
   - Use bullet points with the exact format:
     - [Complete sentence]

2. "Suggested actions to address the smell:"
   - Provide concrete, actionable recommendations to improve documentation completeness.
   - Use a TABLE format with the following column:
     - Action (must follow the exact format: "[LAZY] - [action title]")

Mandatory Output Structure:
- Use only the specified bullet format for evidence.
- Return a VALID JSON object containing the explanations for each affected endpoint.
- Do NOT include any text outside the JSON structure.

input_variables=["lazy_examples", "openapi_json"]
\end{lstlisting}
\end{tcolorbox}

\section{Example of OpenAPI JSON}
\label{app:openapijson-example}

\begin{tcolorbox}[
  title={Box B.1: OpenAPI Specification Excerpt Used for Analysis},
  colback=white,
  colframe=black,
  boxrule=0.6pt,
]
\begin{lstlisting}[style=promptstyle]
{
  "openapi": "3.0.0",
  "info": {
    "title": "Order Management API",
    "description": "API for managing customer orders in an e-commerce platform",
    "version": "1.2.0"
  },
  "paths": {
    "/orders/createNewOrder": {
      "get": {
        "summary": "Create order",
        "operationId": "createNewOrder",
        "requestBody": {
          "required": true,
          "content": {
            "application/json": {
              "schema": {
                "$ref": "#/components/schemas/CreateOrderRequest"  } } } },
        "responses": {
          "200": {
            "description": "Success",
            "content": {
              "application/json": {
                "schema": {
                  "$ref": "#/components/schemas/GenericResponse"
                } } } } } } } },
  "components": {
    "schemas": {
      "CreateOrderRequest": {
        "type": "object",
        "properties": {
          "customer_id": { "type": "string" },
          "email": {    "type": "string"  },
          "phone": {    "type": "string"  },
          "payment_details": {  "type": "object",
            "properties": {
              "card_number": {
                "type": "string",
                "example": "4111111111111111"
              },
              "card_holder": { "type": "string"
              } }} 
      },
        "required": ["customer_id", "email"]
      },
      "GenericResponse": {
        "type": "object",
        "properties": {
          "status": {
            "type": "string",
            "example": "success"
          },
          "data": {
            "type": "object"
          }
        }
      }
   }
  }
}
\end{lstlisting}
\end{tcolorbox}

\section{Example of An Analysis Report}
\label{app:report-example}

\begin{tcolorbox}[
  title={Box B.2: Example of a Hermes Diagnostic Report},
  colback=white,
  colframe=black,
  boxrule=0.6pt
]
\begin{lstlisting}[style=promptstyle]
## API Info
- Title: "Order Management API"

## Endpoint Info
- Method: "GET"
- Path: "/orders/createNewOrder"

## Model
- gpt-oss:120b

## Identified Smells
Lazy, Security, Input, Response, Path & Method

### Explanations

### Lazy
- The summary is vague and the endpoint has no description, providing insufficient guidance about purpose, behavior, or usage.

### Security
- No authentication or authorization requirements are documented, and no security scheme is defined in the specification.

### Input
- Input fields lack semantic clarification and validation constraints such as format, pattern, or complete required declarations.

### Response
- The endpoint uses HTTP 200 for resource creation and provides only a generic response schema, without documented error cases.

### Path & Method
- The path is action-oriented ("createNewOrder"), and GET is used for an operation that semantically represents resource creation.

## Improvement Suggestions

[LAZY] - Improve documentation |
Provide a complete description including endpoint purpose, expected inputs, and outputs, with usage examples.

[SECURITY] - Define authentication |
Add Bearer token requirements and document the corresponding security scheme.

[INPUT] - Add validation |
Specify formats (e.g., UUID, email), validation constraints, and complete required field declarations.

[RESPONSE] - Improve responses |
Use 201 Created for successful creation and define detailed success and error response schemas.

[PATH_AND_METHOD] - Fix REST design |
Replace GET /orders/createNewOrder with POST /orders and adopt a resource-oriented path.

\end{lstlisting}
\end{tcolorbox}

%% file: reference.bib
@article{mastouri2025making, title={Making rest apis agent-ready: From openapi to mcp servers for tool-augmented llms}, author={Mastouri, Meriem and Ksontini, Emna and Kessentini, Wael}, journal={arXiv preprint arXiv:2507.16044}, year={2025} }

@article{bandlamudi2025framework, title={A framework for testing and adapting rest apis as llm tools}, author={Bandlamudi, Jayachandu and Chaudhuri, Ritwik and Gantayat, Neelamadhav and Ghosh, Sambit and Mukherjee, Kushal and Agarwal, Prerna and Sindhgatta, Renuka and Mehta, Sameep}, journal={arXiv preprint arXiv:2504.15546}, year={2025} }

@inproceedings{coblenz2023qualitative,
  title={A Qualitative Study of REST API Design and Specification Practices},
  author={Coblenz, Michael and Guo, Wentao and Voozhian, Kamatchi and Foster, Jeffrey S.},
  booktitle={2023 IEEE Symposium on Visual Languages and Human-Centric Computing (VL/HCC)},
  pages={148--156},
  year={2023},
  organization={IEEE}
}

@article{ni2025doc2agent, title={Doc2Agent: Scalable Generation of Tool-Using Agents from API Documentation}, author={Ni, Xinyi and Jian, Haonan and Wang, Qiuyang and Shah, Vedanshi Chetan and Hong, Pengyu}, journal={arXiv preprint arXiv:2506.19998}, year={2025} }

@article{li2025we,
  title={We urgently need privilege management in mcp: A measurement of api usage in mcp ecosystems},
  author={Li, Zhihao and Li, Kun and Ma, Boyang and Xu, Minghui and Zhang, Yue and Cheng, Xiuzhen},
  journal={arXiv preprint arXiv:2507.06250},
  year={2025}
}

@article{runeson2009case,
  title     = {Guidelines for Conducting and Reporting Case Study Research in Software Engineering},
  author    = {Runeson, Per and H{\"o}st, Martin},
  journal   = {Empirical Software Engineering},
  volume    = {14},
  number    = {2},
  pages     = {131--164},
  year      = {2009},
  publisher = {Springer},
  doi       = {10.1007/s10664-008-9102-8}
}

@inproceedings{casas2021uses,
  title={Uses and applications of the OpenAPI/Swagger specification: a systematic mapping of the literature},
  author={Casas, Sandra and Cruz, Diana and Vidal, Graciela and Constanzo, Marcela},
  booktitle={2021 40th International Conference of the Chilean Computer Science Society (SCCC)},
  pages={1--8},
  year={2021},
  organization={IEEE}
}

@article{panichella2022test,
  title={Test smells 20 years later: detectability, validity, and reliability},
  author={Panichella, Annibale and Panichella, Sebastiano and Fraser, Gordon and Sawant, Anand Ashok and Hellendoorn, Vincent J},
  journal={Empirical Software Engineering},
  volume={27},
  number={7},
  pages={170},
  year={2022},
  publisher={Springer}
}

@article{guo2025mcp,
  title={MCP-AgentBench: Evaluating Real-World Language Agent Performance with MCP-Mediated Tools},
  author={Guo, Zikang and Xu, Benfeng and Zhu, Chiwei and Hong, Wentao and Wang, Xiaorui and Mao, Zhendong},
  journal={arXiv preprint arXiv:2509.09734},
  year={2025}
}

@article{gowda2024best,
  title={Best Practices in REST API Design for Enhanced Scalability and Security},
  author={Gowda, P and Gowda, AN},
  journal={Journal of Artificial Intelligence, Machine Learning and Data Science},
  volume={2},
  number={1},
  pages={827--830},
  year={2024}
}

@misc{openapi20242,
  title        = {OpenAPI Specification},
  author       = {{OpenAPI Initiative}},
  year         = {2024},
  howpublished = {\url{https://spec.openapis.org/oas/latest.html}}
}

@book{masse2011rest,
  title={REST API design rulebook: designing consistent RESTful web service interfaces},
  author={Masse, Mark},
  year={2011},
  publisher={" O'Reilly Media, Inc."}
}

@INPROCEEDINGS {lima2025ictai,
author = { Lima, Rayfran Rocha and de Brito, Irineu Evangelista Cruz and da Cunha Nascimento, Jardel and Nogueira, Elias Nascimento },
booktitle = { 2025 IEEE 37th International Conference on Tools with Artificial Intelligence (ICTAI) },
title = {{ Empowering Conversational Systems Through AI Agents Specialized in Reusable APIs Across Software Ecosystems }},
year = {2025},
volume = {},
ISSN = {},
pages = {1316-1323},
doi = {10.1109/ICTAI66417.2025.00191},
url = {https://doi.ieeecomputersociety.org/10.1109/ICTAI66417.2025.00191},
publisher = {IEEE Computer Society},
address = {Los Alamitos, CA, USA},
month =Nov}

@phdthesis{fielding2000rest,
  title  = {Architectural Styles and the Design of Network-based Software Architectures},
  author = {Fielding, Roy Thomas},
  year   = {2000},
  school = {University of California, Irvine}
}

@article{pautasso2014rest,
  title   = {RESTful Web Services: Principles, Patterns, Emerging Technologies},
  author  = {Pautasso, Cesare and Zimmermann, Olaf and Leymann, Frank},
  journal = {IEEE Software},
  volume  = {31},
  number  = {3},
  pages   = {54--61},
  year    = {2014}
}

@article{robillard2017api,
  title   = {What makes APIs hard to learn? Answers from developers},
  author  = {Robillard, Martin P.},
  journal = {IEEE Software},
  volume  = {34},
  number  = {6},
  pages   = {27--34},
  year    = {2017}
}

@article{araujo2022openapi,
  title={Empirical Analysis on OpenAPI Topic Exploration and Discovery to Support the Developer Community},
  author={da Rocha Araújo, Leonardo and Rodríguez, Guillermo and Vidal, Santiago and Marcos, Claudia and Pereira dos Santos, Rodrigo},
  journal={Computing and Informatics},
  volume={40},
  number={6},
  pages={1345--1369},
  year={2022},
  doi={10.31577/cai_2021_6_1345},
  url={https://www.cai.sk/ojs/index.php/cai/article/view/2021_6_1345}
}

@article{theunissen2022mapping,
  title        = {A mapping study on documentation in Continuous Software Development},
  author       = {Theunissen, Theo and van Heesch, Uwe and Avgeriou, Paris},
  journal      = {Information and Software Technology},
  volume       = {142},
  pages        = {106733},
  year         = {2022},
  doi          = {10.1016/j.infsof.2021.106733},
  url          = {https://doi.org/10.1016/j.infsof.2021.106733}
}

@article{khan2021docs,
  title   = {Automatic Detection of Five API Documentation Smells: Practitioners' Perspectives},
  author  = {Khan, Junaed Younus and Khondaker, Md. Tawkat Islam and Uddin, Gias and Iqbal, Anindya},
  journal = {arXiv preprint arXiv:2102.08486},
  year    = {2021},
  url     = {https://arxiv.org/abs/2102.08486}
}

@article{alzahrani2024software,
  title={Software Systems Documentation: A Systematic Review.},
  author={Alzahrani, Abdullah AH},
  journal={International Journal of Advanced Computer Science \& Applications},
  volume={15},
  number={8},
  year={2024}
}

@inproceedings{du2024evaluating,
  title={Evaluating large language models in class-level code generation},
  author={Du, Xueying and Liu, Mingwei and Wang, Kaixin and Wang, Hanlin and Liu, Junwei and Chen, Yixuan and Feng, Jiayi and Sha, Chaofeng and Peng, Xin and Lou, Yiling},
  booktitle={Proceedings of the IEEE/ACM 46th International Conference on Software Engineering},
  pages={1--13},
  year={2024}
}

@article{white2023prompt,
  title   = {A Prompt Pattern Catalog to Enhance Prompt Engineering with ChatGPT},
  author  = {White, Jules and others},
  journal = {arXiv preprint arXiv:2302.11382},
  year    = {2023}
}

@misc{anthropic2024mcp,
  title        = {Model Context Protocol (MCP)},
  author       = {{Anthropic}},
  year         = {2024},
  howpublished = {\url{https://modelcontextprotocol.io}}
}
